# Design of Battery management system for Residential applications

Poushali Pal[1], Devabalaji K. R[2], S. Priyadarshini[3]

*Research scholar, Department of Electrical and Electronics Engineering, Hindustan Institute of Technology and Science, Padur, Chennai, India*

*Assistant professor, Department of Electrical and Electronics Engineering, Hindustan Institute of Technology and Science, Padur, Chennai, India*

*Assistant professor, Department of Electrical and Electronics Engineering, Hindustan Institute of Technology and Science, Padur, Chennai, India*

***Abstract*** — *Battery management system plays an important role for modern battery-powered application such as Electric vehicles, portable electronic equipment and storage for renewable energy sources. It also increases the life-cycle of the battery, battery state and efficiency. Monitoring the state of charge of the battery is a crucial factor for battery management system. This paper deals with monitoring the state of charge of the battery along with temperature, current for Solar panel fitted with battery for residential application. Microcontroller is used for controlling purpose, analog sensors are used for sensing the parameters of voltage, current. The information of the battery is given with tabular form and shown in photograph. Battery parameters are displayed with the LCD screen.*

## I. INTRODUCTION

Due to the risk of environmental pollution there is a rapid penetration of renewable energy sources in power system. Since the electricity demand is unpredictable, so the intermittent generation of renewable energy will affect the system operation and security. There is a need of storage system to compensate the variation of power generation due to different environmental condition such as wind velocity, solar irradiation and climatic changes The battery backup capacity should be high for standalone power generation. Batteries are used as the storage system for renewable technologies. The performance of the battery relies on the chemical reactions within it. The most commonly used battery for solar PV application is lead-acid battery. Ni-Cd or Ni-Metal hydride batteries are used for portable applications. Battery management system (BMS) is required to maintain the battery operationality, protect the battery from damage and to maintain the balance between generation and demand. Moreover, the number of life cycle of battery can be increased using battery management system. A concise, understandable overview of existing methods, key issues, technical challenges, and future trends of the battery state estimation domain is presented [1]. A smart energy management system is developed for multi-cell batteries for designing a self-healing circuit system, modelling the battery aging process and to improve the effectiveness of battery monitoring [2]. The hardware aspects of battery management systems (BMS) for electric vehicle and stationary applications is focussed in this paper [3]. A review is presented for state of health estimation of battery for driving safely and avoid potential failure for Electric Vehicle [4]. Two approaches were proposed, redundant battery management scheme and effective battery charging optimization for efficient use of Electric Vehicle [5]. A detailed discussion of rules-based management algorithms, optimization-based management algorithms and intelligent-based management algorithms is conferred in this paper [6]. A preliminary list of drivers, barriers, and enablers to end-of-life management of solar panels and battery energy storage systems obtained from a scientific literature review is evaluated. A replacement energy management strategy is conferred to coordinate efficiently hybrid energy storage system supported pumped hydro storage with batteries [7]. Efficient battery management system and optimization techniques of battery charging system for electric vehicles and hybrid energy storage system is developed [8-12]. The battery performance is dependent on the accuracy of state of charge determination of the battery. The SOC estimation methods is dependent on different parameters such as current, voltage, charging-discharging method. Terminal voltage method is based on the measurement of voltage drops during the battery discharging [13]. In Ampere-hour counting method the integral of battery current is having accumulated error. The implementation time is required more for the Discharge test method [14]. Open circuit voltage method is based on the measurement of battery voltage when it is disconnected from the load [14]. SOC and battery internal resistance are varying conversely with each other [12]. Kalman filter is primarily the SOC estimation technique for lithium-ion batteries. The major disadvantage of this method is the complexity in its calculation. Neural network method works on the principle of estimation methodology with discharging current, temperature of the battery, output voltage under different charging condition [12,16]. However most of the methods of battery management are used for electric vehicle application and have neglected the measurement of battery temperature with SOC. To bridge the research gaps in the literature this paper proposes the SOC determination by measuring the terminal voltage along with its temperature, charging-discharging current. Battery discharge behaviour is conventionally monitored by measuring the voltage discharge with respect to time. This method is suitable for all operating conditions of the battery. In the first stage the battery SOC is modelled using terminal voltage method. In the second stage the battery management Simulink model is presented using battery SOC. The result of the Simulink model is presented in the last section with experimental setup.

## II. THEORITICAL MODELLING OF BATTERY MANAGEMENT BASED ON SOC

A test system is considered that consist of two solar panels, two batteries, two buck convertor, two D.C loads and one small controller. A.C loads also can be connected in this system using inverter as shown in the Figure1.

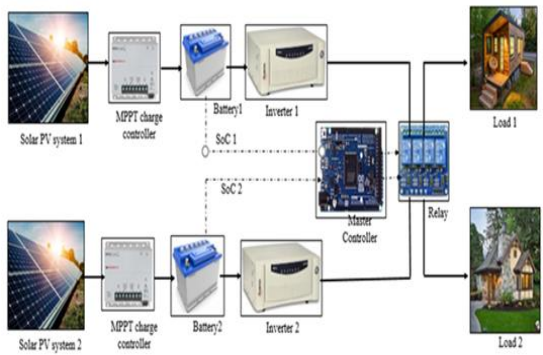

Fig.1.Schematic diagram of battery sharing in Virtual power plant

## III. MODELLING OF BATTERY SOC

$$V_R = I_D * R .........(1)$$

Where $V_R$ =terminal voltage drop by connecting the load

$I_D$ =discharging current

The discharging current of the battery can be measured using Coulomb-counting method. The equation can be formulated as

$$SOC_{t1} = SOC_{t0} + \int_{t0}^{t1} I_D(t)dt / B_{nom} * 100.........(2)$$

Where $SOC_{t1}$ =State of charge of the battery at time $t_1$

Where $SOC_{t0}$ =State of charge of battery at time $t_2$

$B_{nom}$ =Nominal capacity of the battery

$I_D$ =Discharge current of the battery

From these two equations the average discharge current can also be calculated as

$$I_{avg} = [(SOC_1 - SOC_2) * B_{nom}/100 * (1/t_2 - t_1) ........(3)$$

So, the average discharge current can be calculated using this formula. Some experimental results of battery voltage, current and temperature is shown in Table1.

**TABLE1**

**EXPERIMENTAL RESULTS OF BATTERY SOC**

| Time in hour | Voltage | Current | Temperature |
|---|---|---|---|
| 9 h | 26.9 volt | 24.8 A | 62⁰C |
| 10 h | 29.1 volt | 23.2 A | 66⁰C |
| 11 h | 29.0 volt | 21.1 A | 67⁰C |
| 12 h | 27.8 volt | 19.6 A | 66⁰C |
| 1 h | 27.3 volt | 19.4 A | 64⁰C |
| 2 h | 24.1 volt | 17.9 A | 66⁰C |

## IV.MATLAB SIMULINK

The Simulink model for battery management system of Virtual power plant is shown in Fig.2.Four different conditions have considered for battery management system. The DC load is connected to the battery directly. Here comes the implementation of energy management system in which if any one solar panel system is run out of power, instead of taking the power from national grid the respective load can be to supplied from other solar panel system which has sufficient power to supply other load without disturbing the load whereas the sharing system may supply one load at a time. The constitution of loads is shown in Table 2 and Table3.

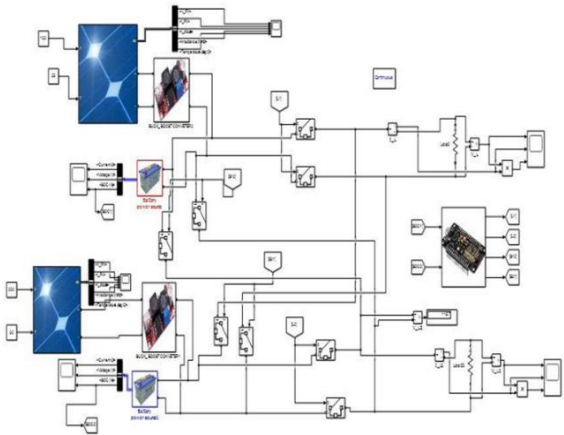

Fig.2 Simulink model for battery management system

Constitution of load 1

Table 2

| Sl.no. | Type of the load | Power consumption of each load in watt | Total power consumption in watt |
|---|---|---|---|
| 1 | Tube light | 45 | 45*6=270 watt |
| 2 | Ceiling Fan | 80 | 80*7=560 watt |
| 3 | Computer | 70 | 100*1=100 watt |
| Total power consumption | | | 930 watt |

Constitution of load 2

Table 3

| Sl.no. | Type of the load | Power consumption of each load in watt | Total power consumption in watts |
|---|---|---|---|
| 1 | Tube light | 45 | 45*6=270 watt |
| 2 | Bulb | 100 | 100*4=400watt |
| 3 | LCD TV | 100 | 100*1=200 watt |
| Total power consumption | | | 870 watt |

The sharing conditions are discussed below with 4 different scenarios:

1. When load 1 is ON, load 2 is OFF, then sharing switch S21 should be ON. (if Solar panel system 2 is sufficient to supply another load)

2. When load 1 is OFF, load 2 is OFF, then sharing switch 1 and 2 should be OFF. (if both the solar panel systems are not sufficient to supply power).

3. When load 1 is ON, load 2 is ON, then sharing switch 1 and sharing switch 2 can be OFF. (Because sharing cannot be done if its own load is working)

4. When load 2 is ON, Load 1 is OFF, then sharing switch S12 should be ON. (if Solar panel system 1 is sufficient to supply another load).

## V. RESULTS AND DISCUSSIONS
### Scenario 1

When the State of charge of the battery 1 is more than 50 % and state of charge of battery 2 is less than 50 %, than the switch 1 should be ON, load 2 should be ON and load 1 should be in OFF condition. The SOC of the batteries is shown in Fig.3. The power consumption of the two loads in watts is shown with Fig.4.

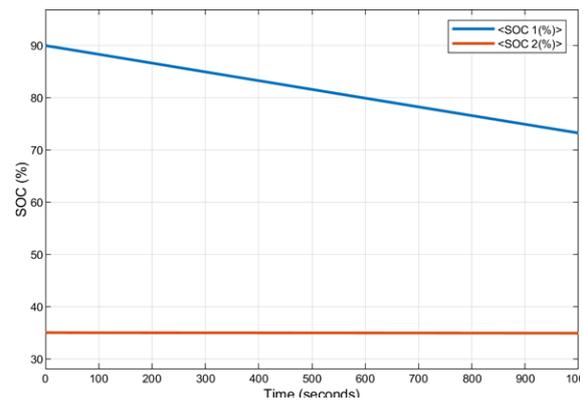

Fig.3 The state of charge of two batteries

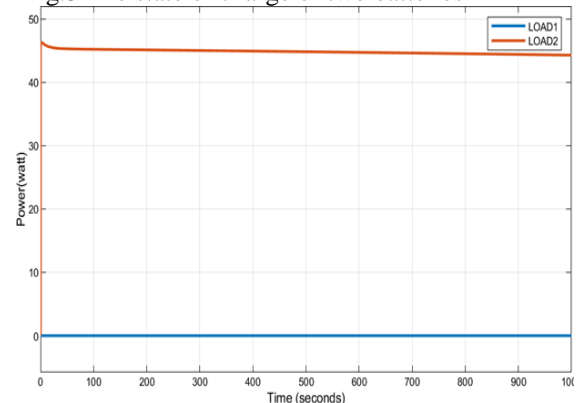

Fig.4 The power consumption for the two loads

Scenario 2

When the state of charge of the two batteries is greater than and equal to 50%, than both the switches should be in OFF condition and both the loads should be in ON condition. The SOC of the batteries is shown in Fig.5 and the power consumption of the two loads is shown in Fig.6.

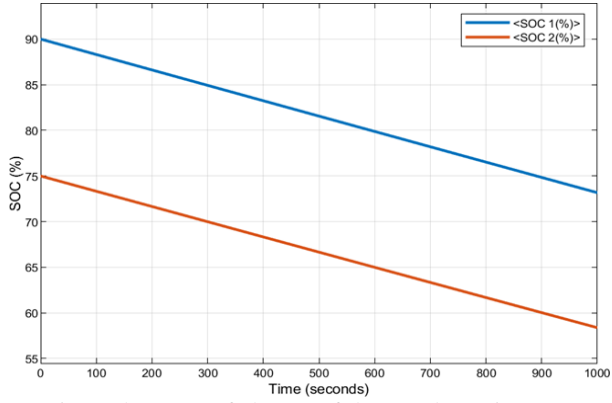

Fig.5 The state of charge of the two batteries

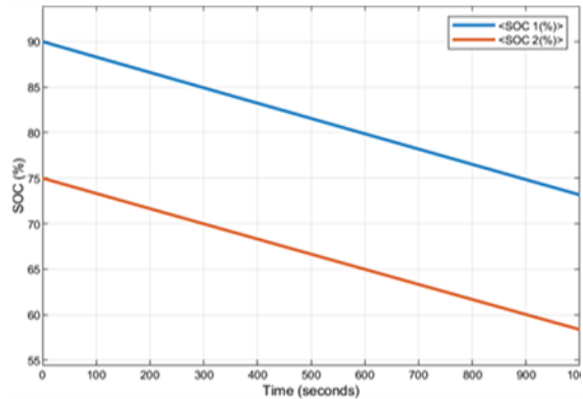

Fig.6 Power consumption of two loads

Scenario 3

When the state of charge of both the batteries is less than 50%, than both the switches should be in OFF condition and both the loads should be in OFF condition. The SOC of the batteries is shown in Fig.7 and the power consumption of the two loads is shown in Fig.8.

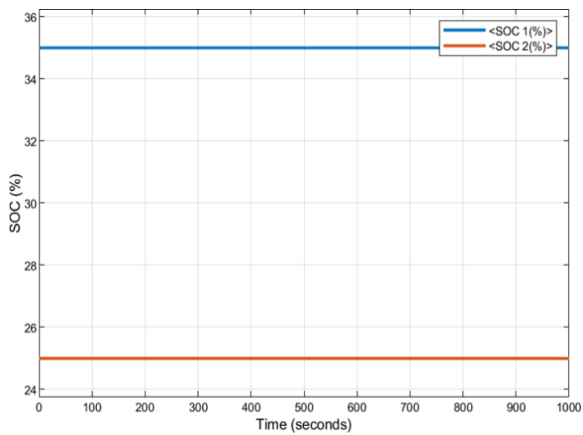

Fig.7 The state of charge of the two batteries

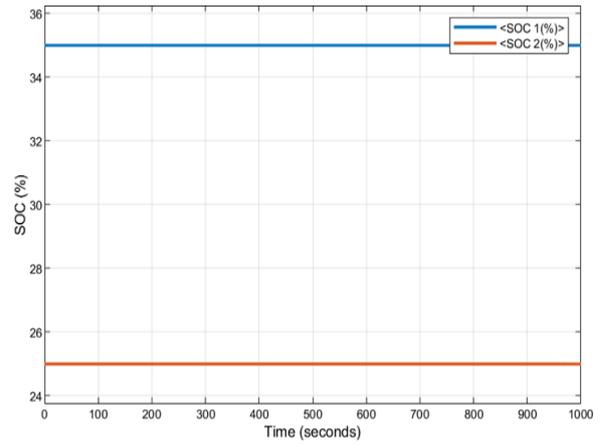

Fig.8 The power consumption of the two loads

Scenario 4

When the state of charge of the battery 1 is less than 50% and the state of charge of the battery 2 is more than 50, than switch 2 and load 1 should be in ON condition and switch 1, load 2 should be in OFF condition is shown in Fig.9 and the power consumption of the two loads is shown in Fig.10.

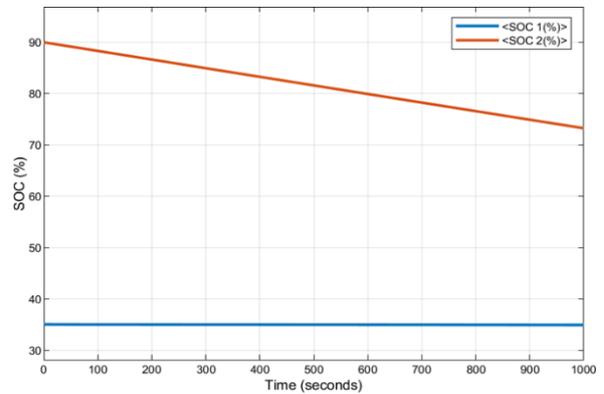

Fig.9 The state of charge of the two batteries

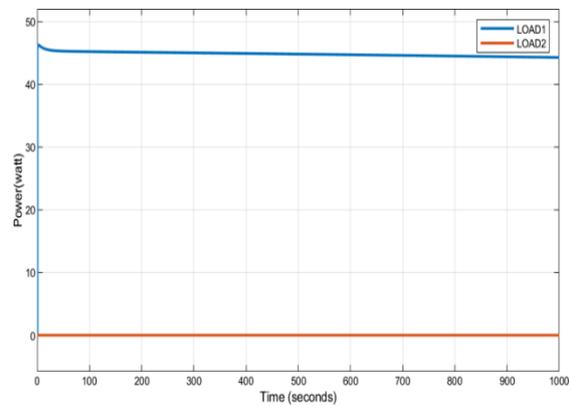

Fig.10 The power consumption of the two loads

The comparison of the four scenarios is shown in Table.4. It is clear from the above scenarios that sharing of the sources can happen when battery SOC is more than 50% and the loads can be supplied accordingly.

Table.4 Comparison between four scenarios

| Sl.no | Percentage SOC of the batteries | Power consumption in watts | Advantage |
|---|---|---|---|
| Scenario1 | SOC1=52%, SOC2=35% | L1=0, L2=35 | No supply to the load 1 |
| Scenario2 | SOC1=90%, SOC2=75% | L1=45 L2=45 | Maximum supply to the loads |
| Scenario3 | SOC1=35%, SOC2=25% | L1=0 L2=0 | No supply to the loads |
| Scenario4 | SOC2=90%, SOC1=35% | L1=45 L2=0 | No supply to the load 2 |

## VI. EXPERIMENTAL SETUP

A hardware system is designed for monitoring the battery SOC. The battery management system will be operated after monitoring SOC and according to that the microcontroller will send the signal to the relay switches. Relay controls the load by opening and closing contacts in the circuit based on the SOC level. The experimental setup for the MPPT solar charge controller is shown in Fig.11.

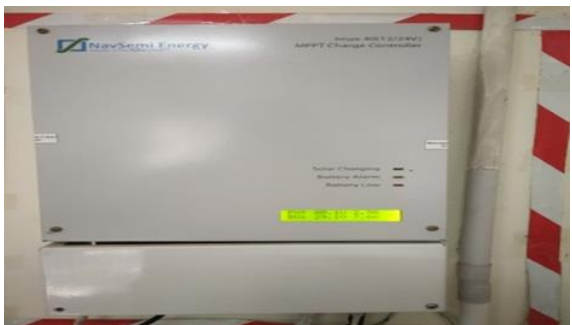

Fig.11 The experimental setup for charge controller

The 12v,200 Ah Lead-acid battery setup with inverter is shown in the Figure.12. The whole system is connected with rooftop solar panel.

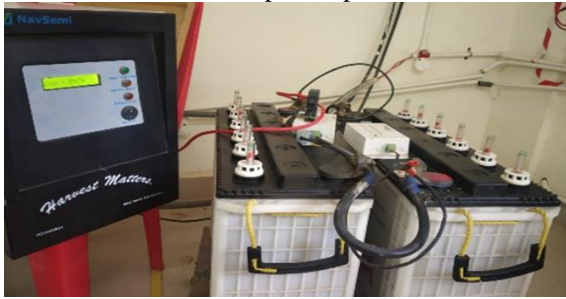

Fig.12 Battery setup

The parameters of the inverter are shown in the Fig13. The battery and panel parameters are shown in the Fig.14.

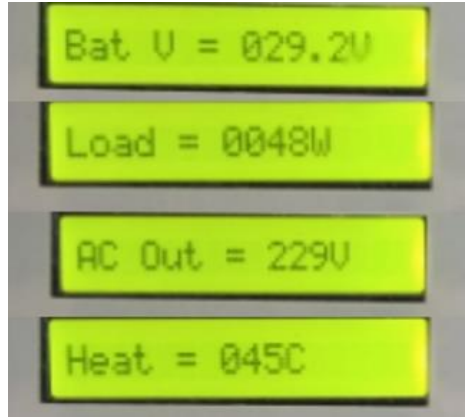

Fig.13 Display of inverter input voltage, load, inverter output voltage and temperature

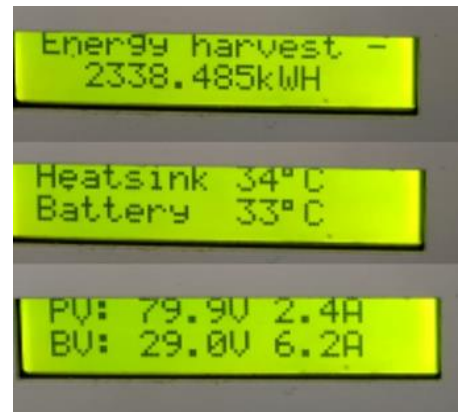

Fig.14 Display of panel voltage, current, Battery voltage, current, battery temperature

## VII. CONCLUSION

The research issues related to the existing method of battery management system is discussed. Based on the different study, the terminal voltage method is used for SOC monitoring. The SOC values are implemented for battery management system in an experimental setup, which can be used for residential application. The researchers will get an idea of efficient battery monitoring system with advance and efficient methodologies of voltage, current and temperature sensing which will improve
• Battery life-cycle and efficiency
• System efficiency

### ACKNOWLEDGMENT

The authors are thankful to Hindustan Institute of Technology and Science for providing the necessary facilities and equipment to implement the novel topic of battery sharing or management estimating battery SOC.